# Transition metal dichalcogenide nanospheres for high-refractive-index nanophotonics and biomedical theranostics

*G.I. Tselikov[†],[1] G.A. Ermolaev[†],[1] A. A. Popov,[2] G.V. Tikhonowski,[2] A.S. Taradin,[1] A.A. Vyshnevyy,[1] A.V. Syuy,[1] S.M. Klimentov,[2] S.M. Novikov,[1] A.B. Evlyukhin,[3,1] A.V. Kabashin,[2,4] A.V. Arsenin,[1,5] K.S. Novoselov,[6,7,8] and V.S. Volkov,[1,5]\**

[1]Center for Photonics and 2D Materials, Moscow Institute of Physics and Technology, Dolgoprudny 141700, Russia.

[2]MEPhI, Institute of Engineering Physics for Biomedicine (PhysBio), Bio-nanophotonics Laboratory, Moscow 115409, Russia.

[3]Institute of Quantum Optics, Liebniz University Hannover, Hannover 30167, Germany.

[4]Aix-Marseille Université, CNRS, LP3, Campus de Luminy, Case 917, Marseille 13288, France.

[5]GrapheneTek, Skolkovo Innovation Center, Moscow 143026, Russia.

[6]National Graphene Institute (NGI), University of Manchester, Manchester M13 9PL, UK.

[7]Department of Materials Science and Engineering, National University of Singapore, Singapore 117574, Singapore.

[8]Chongqing 2D Materials Institute, Liangjiang New Area, Chongqing 400714, China.

*e-mail: volkov.vs@mipt.ru

**ABSTRACT**

Recent developments in the area of resonant dielectric nanostructures has created attractive opportunities for the concentrating and manipulating light at the nanoscale and the establishment of new exciting field of all-dielectric nanophotonics. Transition metal dichalcogenides (TMDCs) with nanopatterned surfaces are especially promising for these tasks. Still, the fabrication of these structures requires sophisticated lithographic processes, drastically complicating application prospects. To bridge this gap and broaden the application scope of TMDC nanomaterials, we report here femtosecond laser-ablative fabrication of water-dispersed spherical TMDC ($MoS_2$ and $WS_2$) nanoparticles (NPs) of variable size (5 – 250 nm). Such nanoparticles demonstrate exciting optical and electronic properties inherited from TMDC crystals, due to preserved crystalline structure, which offers a unique combination of pronounced excitonic response and high refractive index value, making possible a strong concentration of electromagnetic field in the nanoparticles. Furthermore, such nanoparticles offer additional tunability due to hybridization between the Mie and excitonic resonances. Such properties bring to life a number of nontrivial effects, including enhanced photoabsorption and photothermal conversion. As an illustration, we demonstrate that the nanoparticles exhibit a very strong photothermal response, much exceeding that of conventional dielectric nanoresonators based on Si. Being in a mobile colloidal state and exhibiting superior optical properties compared to other dielectric resonant structures, the synthesized TMDC nanoparticles offer opportunities for the development of next-generation nanophotonic and nanotheranostic platforms, including photothermal therapy and multimodal bioimaging.

**INTRODUCTION**

The interest in layered materials[1] has exploded in the last decade, with applications spanning from electronics[2,3] and photonics[4,5] to medicine.[6,7] In particular, the family of transition metal dichalcogenides (TMDCs), most notably $MoS_2$ and $WS_2$, greatly accelerated the progress in compact photodetectors,[8,9] electro- and photocatalysis,[10,11] ultrasensitive detectors,[12,13] and cancer therapies.[14,15] TMDCs exhibit a strong excitonic response, leading to non-trivial optical phenomena enabled by strong light-matter interactions: exciton-polariton transport,[16,17] enhanced second and third harmonic generation,[18,19] high refractive index and giant optical anisotropy.[17,20,21] Motivated by these benefits, novel TMDC nanostructures are being rapidly developed. Typically such nanostructures are produced by electron beam lithography,[22] reactive ion etching,[23] focused ion beam,[24] or laser thinning.[25,26] Despite their effectiveness, all of these methods have a relatively low throughput and create nanostructures formed in place and immobilized on a substrate, significantly limiting their applicability.

Pulsed laser ablation in liquids has recently emerged as a novel pathway to synthesize nanoparticles. In this technique the production of nanoclusters is done through the action of pulsed laser radiation on a solid target, followed by their release into a liquid ambient to form colloidal nanoparticle solutions.[27,28] This method is exceptionally efficient in the case of ultrashort (femtosecond) regime of laser ablation, known as a "fine" regime, which makes possible excellent control of nanoparticle size and the conservation of physicochemical properties of the bulk target material.[29–31] Our previous studies illustrate a high efficiency of fs laser ablation in the fabrication of a variety of nanomaterials exhibiting unique structural and optical properties, including Au,[29,30] Si,[31] TiN,[32] Bi[33] nanoparticles and Au-Si nanocomposites,[34] which promises its successful use for other classes of materials.

Here, we explore the technique of fs laser ablation in water for the synthesis of nanostructures based on TMDCs ($MoS_2$ and $WS_2$). We show that the fs laser-ablative process results in the production of nearly spherical $MoS_2$ and $WS_2$ NPs, ranging from 5 nm to hundreds of nm. We also show that the synthesized NPs conserve crystal structure and, more crucially, high refractive index and excitonic properties of the original crystals. As a result, they exhibit distinct Mie-resonances and provide a significantly improved photothermal performance compared to conventional dielectric nanoresonators in the optical transparency window of biological tissue. The proposed approach opens up the substrate-free nanofabrication of functional TMDC nanostructures, which promises the tremendous potential for nanophotonic and biomedical applications.

## RESULTS AND DISCUSSION

### Laser ablation synthesis of TMDC NPs

For the production of $MoS_2$ and $WS_2$ NPs from bulk crystal (Figure 1a), we adopted the methods of ultra-short pulsed laser ablation[29–31] (Methods) in water, as illustrated in Figure 1b. Within a few minutes of the procedure, we observed a green-brownish coloration of the solution, indicating the formation of TMDC NPs (Figure 1c). Synthesized colloids were extremely stable, showing practically no signs of aggregation or precipitation even after months of storage at room temperature. This remarkable stability of colloidal solutions can be attributed to a significant charge of the NP surface identical to the process found for gold NPs.[30] It should be noted that we employed a high-numerical-aperture objective to increase laser fluence and thus maximize laser ablation efficiency, but such a geometry results in a high fluence gradient and, hence, in a relatively broad size distribution.[28]

We centrifuged colloidal solutions at increasing rotation speeds of 200 – 8000 rpm to achieve monodisperse size distribution and remove the smallest (< 10 nm) NPs (Methods). Then, size distributions of the formed colloids were assessed by the analysis of scanning electron microscopy images of NPs (Methods). As expected, after the centrifugation at high rotation speed, the solutions exhibited a narrow lognormal distribution presented in Figure 1d. In addition, the colloids centrifuged under different conditions had smoothly changing hues (Figure 1c) due to variable size and material's high dielectric permittivity (Figure 1e) acquired by ellipsometry (Methods). Therefore, pronounced colors already indicate that NPs retain optical characteristics of the original crystal. As described in the following sections, we also performed extensive morphological and optical analyses to establish this result quantitatively.

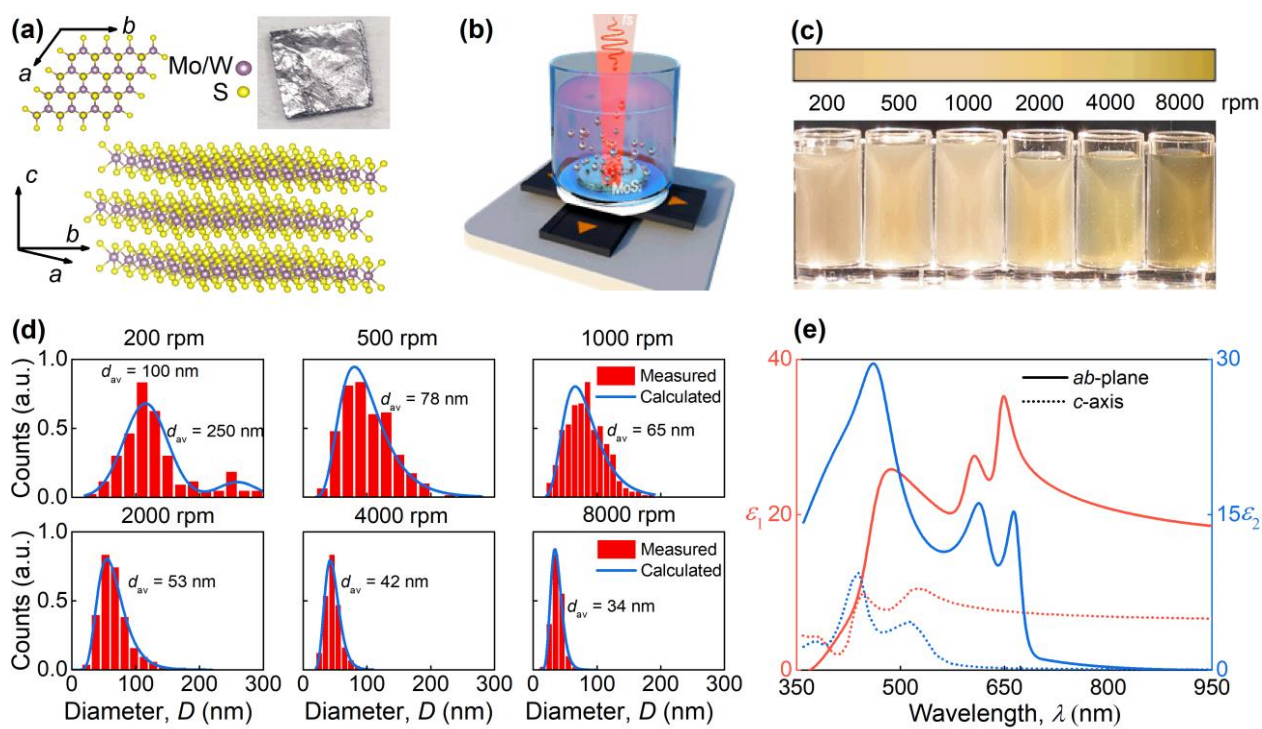

**Figure 1. Engineering of MoS$_2$ NPs. a** Illustration of the MoS$_2$/WS$_2$ layered structure. The inset is the photograph of synthetically synthesized bulk MoS$_2$ crystal. **b** Schematic representation of femtosecond pulsed laser ablation setup used for the fabrication of MoS$_2$ and WS$_2$ colloidal NPs. **c** Photo of colloidal solutions of laser-ablated MoS$_2$ NPs separated by the mean size of nanoparticles by centrifugation at different rotation speeds: 200, 500, 1000, 2000, 4000, and 8000 rpm (from left to right). **d** Size distributions of laser-ablated MoS$_2$ NPs separated by the mean size of nanoparticles by centrifugation at different rotation speed: 200, 500, 1000, 2000, 4000, and 8000 rpm. **e** Real ($\varepsilon_1$) and imaginary ($\varepsilon_2$) parts of the dielectric function along the *ab*-plane and *c*-axis.

**Morphology of TMDC NPs**

Scanning electron microscopy (SEM) (Figure 2a and Supplementary Figure 1a), shows that the produced NPs are almost spherical. The energy dispersive X-ray (EDX) spectrum in the inset of Figure 2a confirms the predominant atomic composition of the original crystal with little oxygen present (less than 5%). Oxygen may have been generated as a result of minor oxidation during laser ablation in water. Meanwhile, selected area electron diffraction (SAED) images (Figure 2b for MoS$_2$ and Supplementary Figure 1b for WS$_2$) show a ring sequence typical for TMDCs with a hexagonal lattice. According to the Joint Committee on Powder Diffraction Standards (card No: 75-1539), the rings labeled 1 – 5 in Figure 2b correspond to the first five diffraction rings of the 2H-MoS$_2$ structure. The calculated lattice constants (*a* = *b* = 0.32 nm and *c* = 1.39 nm) from the rings correspond to lattice values for bulk MoS$_2$ from the literature[17]. Supplementary Table 1 contains the final analysis findings of the SAED pattern. The internal structure of synthesized NPs was then examined using transmission electron microscopy (TEM). TEM data shown in Figure 2c (for MoS$_2$) and Supplementary Figure 1c-d (for WS$_2$) evidence a polycrystalline core structure and a thick fullerene-like outer shell. A closer look reveals that the NPs shell is not fully spherical and has a polygonal shape. This finding suggests that NPs formation takes place along crystallographic axes, similar to the recently observed anisotropic crystallographic etching in TMDC during the lithography process.[23]

At the same time, the smallest NPs (below 10nm in diameter), demonstrate well-developed layered structure and complete absence of the shell, see the inset of Figure 2c. Hence, one could expect them to inherit the original crystal's giant optical anisotropy.[17] As a result, small TMDC NPs may pave the way for a new research field related to extremely anisotropic quantum dots. Finally, Raman spectroscopy was used to characterize the NPs (Figure 2d and Supplementary Figure 2). The characteristic Raman peaks ($E_{2g}^1 = 383$ cm$^{-1}$ and $A_{1g} = 408$ cm$^{-1}$) perfectly matched the values reported for bulk MoS$_2$,[35] additionally demonstrating the excellent quality of produced NPs.

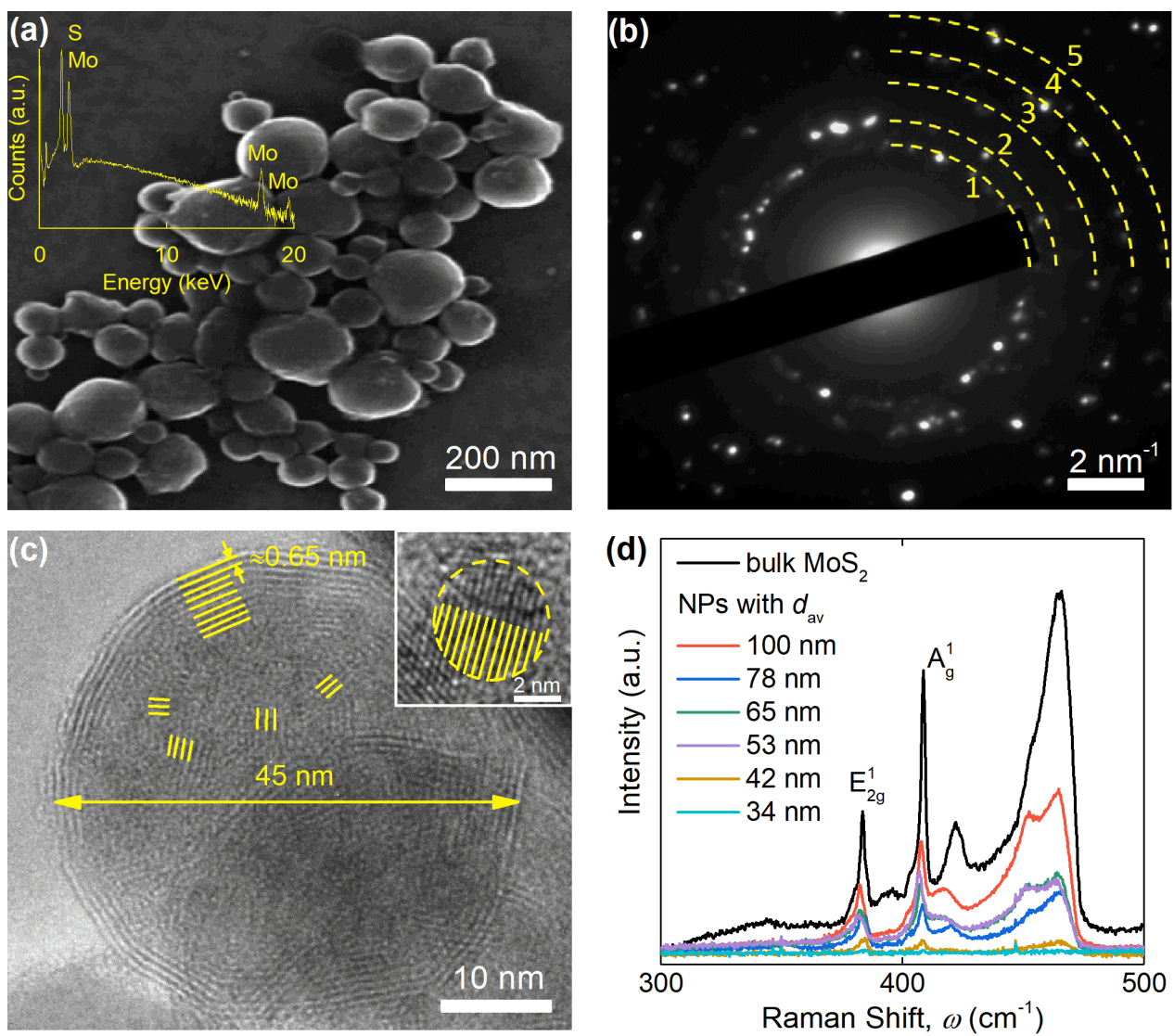

**Figure 2. Structure of MoS$_2$ NPs. a** SEM image of laser-ablated MoS$_2$ nanoparticles selected from a colloidal solution at 200 rpm. Inset shows the EDS spectrum of MoS$_2$ NPs on a Si substrate. **b** SAED pattern for laser-ablated MoS$_2$ NPs. **c** TEM image of 45 nm MoS$_2$ nanoparticle with a ring sequence typical for the hexagonal lattice. Inset shows a monocrystalline inner structure of 5 nm MoS$_2$ NP. Yellow lines highlight crystal planes. **d** Resonant Raman scattering spectra of laser-ablated MoS$_2$ NPs of different average sizes: 100 nm, 78 nm, 65 nm, 53 nm, 42 nm and 34 nm. The excitation wavelength is 633 nm.

**Optical response of TMDC NPs**

In the light of the high crystallinity of synthesized NPs, one can expect them to exhibit bulk high refractive index and excitonic properties. The measured extinction curves (Figure 3a and Supplementary Figure 3) exhibit rich structure that agrees well with the predictions based on anisotropic constants from Figure 1d and the extended Mie-theory for radially anisotropic spheres (Methods).[36] As a result, optical properties of NPs mimic those of the original crystal, which opens up exciting possibilities for all-dielectric nanophotonics due to very high dielectric permittivity ($\varepsilon$ ~ 22) of TMDCs in the visible range, much exceeding that of traditional high-refractive-index materials like Si ($\varepsilon$ ~ 14)[37] and GaP ($\varepsilon$ ~ 10).[38] Since the refractive index determines the capacity of NP to control multipole moments and their ability to concentrate electromagnetic energy, one can expect the excitation of distinct Mie-multipole resonances from TMDCs nanostructures,[37] particularly magnetic (MD) and electric dipole (ED) resonances. Figure 3a clearly shows a blue-shift of the MD resonance as the size of NPs decreases from 100 to 42 nm. Meanwhile, silicon NPs of the same size do not exhibit any features in the 1.5 – 2 eV (620 – 830 nm) spectral range, as follows from Figure 3b. Moreover, MD and ED resonances appear near A- and B-excitons, as illustrated in Figure 3c. Additionally, the high refractive index of $MoS_2$ leads to the excitation of Mie resonances that are substantially red-shifted compared to silicon NPs of the same size (Figure 3d) and match the window of relative biological transparency (700 – 980 nm), which promises attractive theranostics (therapy+diagnostics) applications.

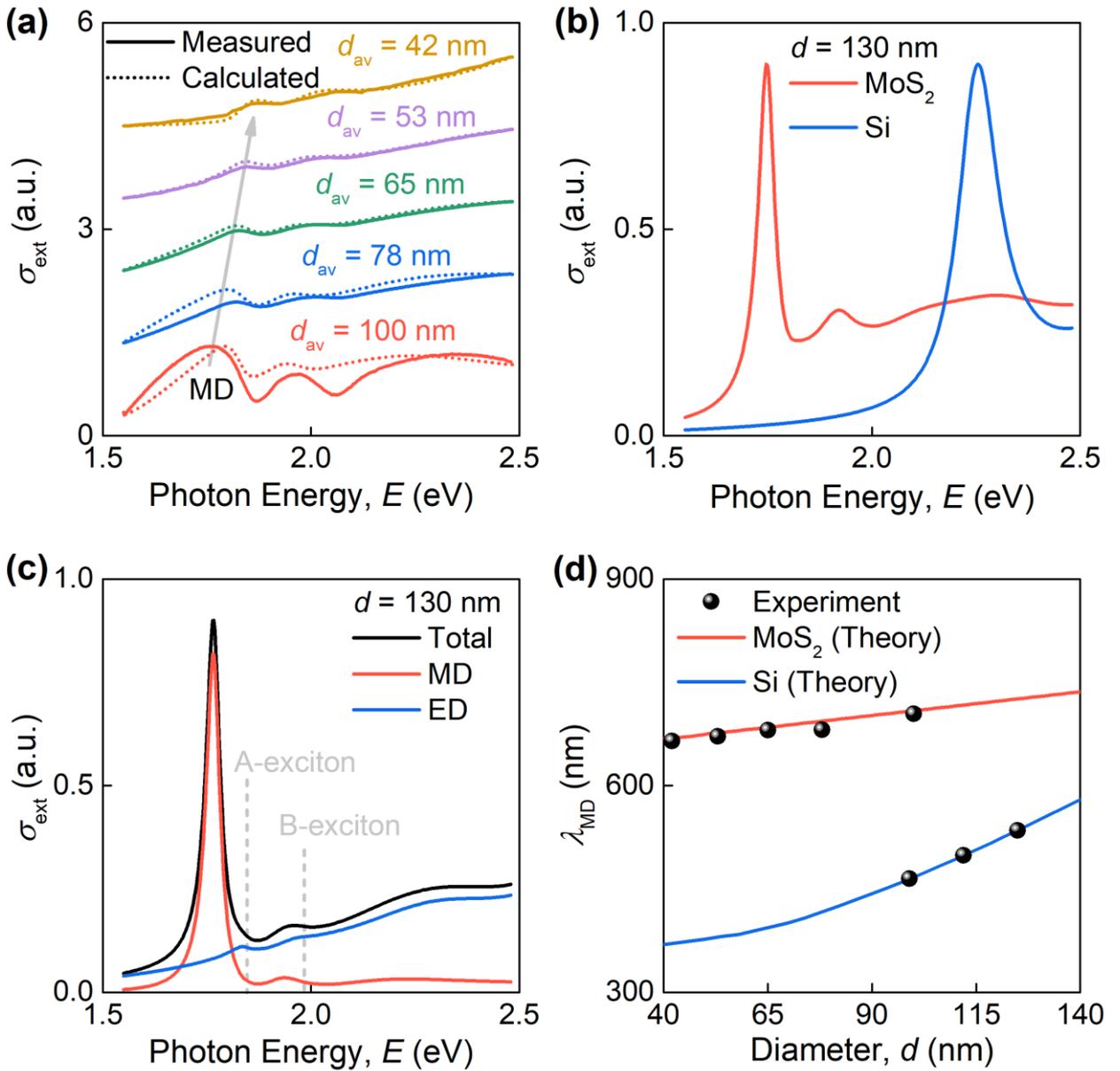

**Figure 3. Optical properties of MoS₂ NPs. a** Extinction cross-section of colloidal MoS$_2$ NPs of different sizes indicated in the graph. Solid and dashed lines show experimental and calculated data, correspondingly. **b** Calculated extinction cross-section of 130 nm MoS$_2$ and Si NPs. **c** Multipole decomposition of extinction spectrum of a 130 nm MoS$_2$ NP. **d** Dependence of magnetic dipole (MD) peak position on the size of MoS$_2$ and Si NPs.

**Photothermal applications of TMDC NPs**

Photothermal therapy of cancer[39] presents one of the most appealing implementations of nanomedicine. However, this therapeutic modality requires efficient nanosensitizers absorbing in the tissue transparency window and having a relatively small size (less than 100 nm) for a better transport in vivo and an enhanced uptake by the cells. To assess the potential of laser-synthesized TMDC NPs for these applications, we investigated power-dependent Raman spectra from 65 nm $MoS_2$ and $WS_2$ NPs (Figure 4a-b and Supplementary Figure 4a-b). To determine heating efficiency, we assumed linear temperature $T$ dependence of characteristic Raman peaks ($E_{2g}^1$ and $A_{1g}$) position $\omega(T)$:

$$\omega(T) = \omega_0 + \chi_1 T$$

where $\omega_0$ is the peak position of modes at room temperature and $\chi_1$ is the first-order temperature coefficient determined in the recent work.[40] Consequently, we were able to recalculate the thermal shift of Raman peaks into the increase of local temperature (Figure 4c-d and Supplementary Figure 4c-d). Surprisingly, the temperature reached almost 800 K even at a very low excitation power of 300 μW, which was obviously due to drastically enhanced NPs absorption boosted by the excitonic transitions. Apart from the 532 and 633 nm excitation wavelength used in Raman, we also tested the photothermal response of $MoS_2$ NPs at 830 nm in the therapeutic window (700 – 980 nm) and compared it with silicon NPs. The concentrations of NPs were normalized to the value of extinction at 830 nm (Figure 4e). Under the same excitation conditions, $MoS_2$ NPs provided almost two times greater photothermal response than silicon NPs (Figure 4f) due to a higher refractive index resulting in superior electromagnetic field concentration.[37] This high photothermal response makes TMDC NPs a perfect biocompatible agents for photothermal applications.

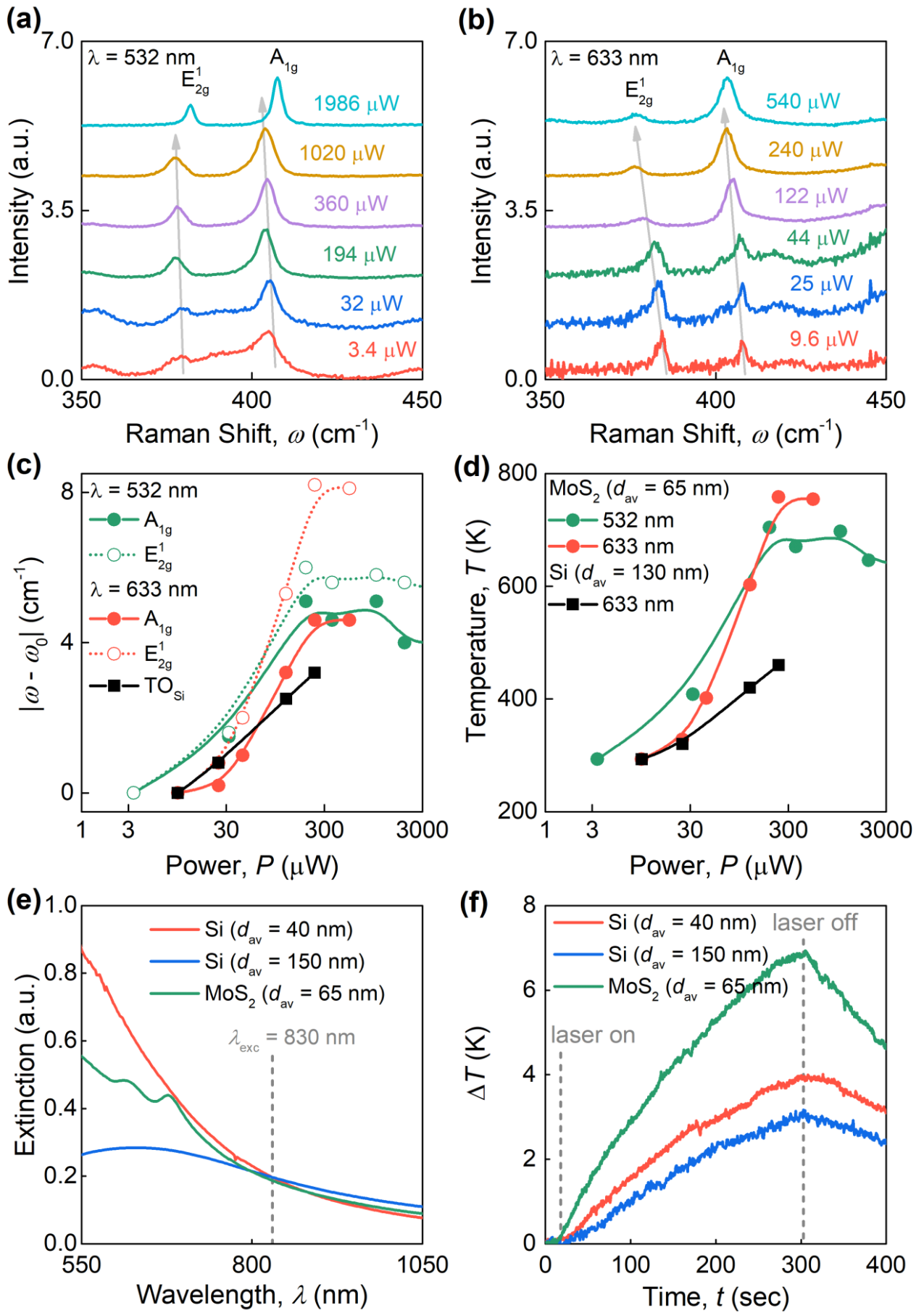

**Figure 4. Raman and photothermal responses of MoS$_2$ NPs.** Raman spectra of 65 nm MoS$_2$ NPs on a glass substrate recorded under different laser excitation powers for two pumping wavelengths of 532 nm **(a)** and 633 nm (**b**). **c** Temperature shift of the frequencies corresponding to $E^1_{2g}$ and $A_{1g}$ Raman scattering modes of 65 nm MoS$_2$ NPs on a glass substrate for pumping wavelengths of 532 nm and 633 nm. For comparison, a shift of Si TO phonon mode for Mie-resonant 130 nm Si NPs under resonant 633 nm excitation is given. **d** Dependence of MoS$_2$ and Si NPs local temperature on the power *P* of excitation radiation for pumping wavelengths of 532 nm and 633 nm. **e** Extinction spectra of laser-ablated MoS$_2$ and Si NPs with different sizes used for photothermal measurements. Concentration of NPs was normalized at the excitation wavelength $\lambda_\text{exc}$ = 830 nm. **f** Temperature growth of colloidal MoS$_2$ and Si solutions versus time during constant photothermal excitation with a power of *P* = 1 W at $\lambda_\text{exc}$ = 830 nm.

In summary, we adopted methods of fs-laser ablation for the synthesis of TMDC spherical nanoparticles of a controllable size and low size dispersion. Surprisingly, obtained NPs preserved the original crystal's layered structure and therefore demonstrated a unique combination of pronounced excitonic response and high refractive index value, which renders possible a strong concentration of electromagnetic field in nanoparticles and the hybridization between the Mie and excitonic resonances. Such unique properties make TMDC nanoparticles very promising candidates for a variety of all-dielectric nanophotonic applications where enhanced light-matter interaction via electromagnetic resonances plays a key role, including nanolasing,[41] biosensing,[42] photothermal cancer therapy,[43] and nonlinear optics.[44] Furthermore, in contrast to conventional lithographic techniques for the fabrication of dielectric resonators, the proposed laser-ablative synthesis is substrate-free, high throughput, and makes possible the fabrication of nanomaterials in mobile (colloidal) state, which promises the extension of TMDC-based materials to medical applications, including cancer therapy[43] and biomedical imaging.[45] Finally, we believe our synthesis approach is universal for all layered materials such as MXenes,[46] TMDCs,[47] graphite,[48] and hyperbolic materials.[49] The fabrication of mobile nanostructures of these materials promises the development of post-silicon nanotechnology and a variety of novel attractive applications.

## ACKNOWLEDGMENTS

We gratefully acknowledge the financial support from the Ministry of Science and Higher Education of the Russian Federation (Agreement No. 075-15-2021-606). Characterization of fabricated solutions of TMDC nanoparticles (G.I.T.) was supported by the Russian Science Foundation (grant № 21-79-00206). Calculation of the extinction spectra (A.A.V.) was supported by the Russian Science Foundation (grant № 20-79-00349). Fabrication of TMDC nanoparticles was supported by the Russian Science Foundation (grant № 19-72-30012). K.S.N acknowledges support from the Ministry of Education (Singapore) through the Research Centre of Excellence program (Award EDUN C-33-18-279-V12, Institute for Functional Intelligent Materials).

## AUTHOR CONTRIBUTIONS

[†]These authors contributed equally. G.I.T., A.V.K., A.V.A., K.S.N., and V.S.V. suggested and directed the project. G.I.T., G.A.E., A.A.P., G.V.T., A.V.S., S.M.K., and S.M.N. performed the measurements and analyzed the data. A.A.V., A.S.T., A.B.E., G.I.T., and G.A.E. provided theoretical support. G.I.T. and G.A.E. wrote the original manuscript. G.I.T., G.A.E., A.A.V., A.B.E., A.V.K., A.V.A., K.S.N., and V.S.V. reviewed and edited the paper. All authors contributed to the discussions and commented on the paper.

## COMPETING INTERESTS

The authors declare no competing interests.

## METHODS

**Sample preparation.** For the fabrication of TMDC NPs, we used a diode pump Teta 10 system (Avesta, Russia) with 100 µJ pulse energy at 1030 nm and a repetition rate of 10 kHz. The laser beam was focused onto the surface of highly oriented synthetic 2H-phase $MoS_2$ and $WS_2$ crystals (2D Semiconductors Inc., USA) placed on the glass vessel bottom filled with deionized water. The process was carried out for 0.5 h at room temperature with a final NP concentration of about 0.1 mg/ml.

**Structural characterization.** The atomic composition of the synthesized NPs was characterized by a scanning transmission electron microscopy (STEM) system (MAIA 3, Tescan, Czech Republic) operating at 0.1–30 kV coupled with an EDS detector (X-act, Oxford Instruments, High Wycombe, UK). Samples for scanning electron microscopy (SEM) imaging were prepared by dropping 2 μL of the NPs solution onto a cleaned silicon substrate with subsequent drying at ambient conditions. Morphological and structural properties of synthesized NPs were characterized by the high-resolution transmission electron microscopy (HR-TEM) system (JEOL JEM 2010) operating at 200 kV with a Gatan Multiscan CCD in imaging and diffraction modes. Samples were prepared by dropping 2 μL of NPs solution onto a carbon-coated TEM copper grid and subsequent drying at ambient conditions. Analysis of selected area electron diffraction (SAED) pattern was performed using ProcessDiffraction v.8.7.1 software.

**Optical characterization.** Spectroscopic ellipsometry measurements were performed on an imaging Accurion nanofilm_ep4 ellipsometer over a wide wavelength range 360 – 1000 nm in steps of 1 nm at multiple incidence angles (50°, 55°, 60°). The optical extinction spectra of colloidal NPs were measured using UV-VIS spectrophotometer (Cary 5000, Agilent technologies) in 500 – 800 nm (1.55 – 2.48 eV) spectral interval with the spectral resolution of 1 nm using 2 mm optical path length cuvettes. Raman scattering spectra were collected from NPs deposited on a glass substrate in backscattering geometry using a confocal scanning Raman microscope Horiba LabRAM HR Evolution (HORIBA Ltd., Kyoto, Japan). All measurements were carried out using linearly polarized excitation at wavelengths 633 nm and 532 nm, 1800 lines/mm diffraction grating, whereas we used unpolarized detection to have a significant signal-to-noise ratio. Exciting radiation was focused on the sample surface with the ×100 objective (N.A. = 0.90) into the spot size of ~0.5 μm.

**Numerical simulation.** We used the extended Mie theory for spherical particles with radial anisotropy for numerical simulations.[36] To achieve high accuracy, multipole expansion was carried out up to hexadecapole terms. Extinction spectra for nanoparticle solutions produced by centrifugation were calculated for spheres with diameters ranging from 6 nm to 500 nm with a step of 2 nm, then averaged with weights corresponding to the fitted size distribution functions (Figure 1e).

**Multipolar mode decomposition.** The multipole mode decomposition of the radiation is performed by applying the approach described previously.[50] The fundamental multipole moments are evaluated by integrating the total electric field induced by a normally incident plane wave numerically inside the nanoparticle

## DATA AVAILABILITY

The datasets generated during and/or analyzed during the current study are available from the corresponding author on reasonable request.

Theranostics. *Adv. Opt. Mater.* 1801728 (2019).

46. Anasori, B., Lukatskaya, M. R. & Gogotsi, Y. 2D metal carbides and nitrides (MXenes) for energy storage. *Nat. Rev. Mater.* **2**, 16098 (2017).

47. Manzeli, S., Ovchinnikov, D., Pasquier, D., Yazyev, O. V. & Kis, A. 2D transition metal dichalcogenides. *Nat. Rev. Mater.* **2**, 17033 (2017).

48. Hernandez, Y. *et al.* High-yield production of graphene by liquid-phase exfoliation of graphite. *Nat. Nanotechnol.* **3**, 563–568 (2008).

49. Ma, W. *et al.* In-plane anisotropic and ultra-low-loss polaritons in a natural van der Waals crystal. *Nature* **562**, 557–562 (2018).

50. Evlyukhin, A. B., Reinhardt, C., Evlyukhin, E. & Chichkov, B. N. Multipole analysis of light scattering by arbitrary-shaped nanoparticles on a plane surface. *J. Opt. Soc. Am. B* **30**, 2589 (2013).